\documentstyle[preprint,eqsecnum,aps,psfig,epsf]{revtex}
\def\ber{\begin{eqnarray}}
\def\eer{\end{eqnarray}}
\def\beq{\begin{equation}}
\def\eeq{\end{equation}}
\def\bra{\langle}
\def\ket{\rangle}
\def\fpi{f_{\pi}}
\def\qpar{\epsilon}

\begin{document}
\baselineskip 0.6cm
\draft
\preprint{SNUTP-97-046}
\title{Generalized Electric Polarizability of the Proton from Skyrme Model}
\author{Myunggyu Kim and Dong-Pil Min}
\address{Center for Theoretical Physics, Seoul National University,
Seoul 151-742, Korea}
\date{\today}
\maketitle
\begin{abstract}
We calculate the electric polarizability $\alpha(q^2)$ of the proton 
in virtual Compton scattering using the Skyrme model. 
The $q^2$ dependence of the polarizability is
comparable with the predictions obtained from the non-relativistic quark model 
and the linear sigma model. 
The chiral behaviors of our $\alpha(0)$ and $d^2\alpha(0)/d^2q^2$ agree with 
the results of the chiral perturbation theory. The discrepancy can be traced
back to the contribution of the intermediate $\Delta$ state  
degenerate with the $N$ which is a characteristic of a large-$N_C$  model.
\end{abstract}

\newpage 
\section{Introduction}
Recently there has been much study on the virtual Compton scattering (VCS).
Several experiments have been proposed and are being performed in order to 
get more information about the structure of the nucleon via the process
~\cite{exp1,exp2,exp3,exp4}. 

At low energies, real Compton scattering from the nucleon are known to be
described in terms of
two quantities for the nucleon structure - the electric ($\alpha$) 
and the magnetic ($\beta$) polarizability. It is a natural thought that 
the virtual Compton scattering could provide even more information on the
structure of the nucleon. They are called the generalized polarizabilities
 and are functions of the four-momentum transfer $q^2$ of the in-going virtual
photon. The virtual photon could also carry a longitudinal polarization.

Theoretically  Guichon, Liu and Thomas developed the formalism 
of the generalized polarizabilities~\cite{exp4}. 
They analyzed the structure-dependent
part beyond the low energy theorem  in terms of a multipole expansion. 
They
also gave a first estimate for the ten generalized polarizabilities using
the non-relativistic quark model and elaborated the study including
the recoil effects~\cite{GLT2}. Using an effective Lagrangian 
Vanderhaeghen~\cite{ET} obtained the $q^2$ behavior of the
electric and the magnetic polarizabilities of the proton. The slopes of
the electric and the magnetic polarizabilities were predicted by
several authors with several frameworks, such as the linear sigma model
~\cite{LSM} and the heavy-baryon formulation of the chiral perturbation
theory~\cite{CHPT1}.

In this letter, we attempt a different point of view  to describe 
the generalized electric polarizability based on a topological
soliton model (Skyrme model). Within many soliton models, only the
expression and the value of $\alpha$ at $q^2=0$ have been obtained.
The present work is the first trial to give the slope of $\alpha$
among the topological and non-topological soliton models.
We also obtain the leading chiral terms of $\alpha$ and its slope
and show that these agree with the results of the linear sigma model and
the heavy-baryon chiral perturbation theory.

\section{The Skyrme model in the presence of electromagnetic fields}
Our starting point is a $U(1)$-gauged effective model.
The gauged Skyrme Lagrangian is
\beq
{\cal L}=\frac{\fpi^2}{4}Tr[D_\mu U^{\dag} D^\mu U]
+\frac{1}{32\qpar^2}Tr\left[ [U^{\dag} D_\mu U,U^{\dag} D_\nu U]^2\right]
+\frac{\fpi^2}{2}m_{\pi}^2(2-Tr[U])
\eeq
where $\fpi$ is the pion decay constant, $\qpar$ is the dimensionless Skyrme
parameter and the $U$ is the $SU(2)$ chiral field. The covariant derivative is
defined as
\beq
D_\mu U=\partial_\mu U+i e A_\mu[Q,U],
\eeq
where $A_\mu$ is the electromagnetic field and Q is the charge matrix.
\beq
Q=\frac{1}{6}+\frac{\tau_3}{2}.
\eeq
The Lagrangian can be decomposed into three terms depending on the number of
the gauge field.
\ber
\label{eq:lagt}
{\cal L}&=&{\cal L}_0+{\cal L}_1+{\cal L}_2, \\
\label{eq:lag0}
{\cal L}_0&=&\frac{\fpi^2}{4}Tr[L_\mu L^\mu]
+\frac{1}{32\qpar^2}Tr\left[ [L_\mu,L_\nu]^2\right] 
+\frac{\fpi^2}{2}m_{\pi}^2(2-Tr[U]), \\
\label{eq:lag1} 
{\cal L}_1&=&e A_\mu J^\mu, \\
\label{eq:lag2}
{\cal L}_2&=&-\frac{1}{2}e^2 A_\mu S^{\mu\nu} A_\nu, \\
J_\mu &=& i\frac{\fpi^2}{2}Tr[Q(L_\mu+R_\mu)]
+i\frac{1}{8\qpar^2}
Tr\left[Q\left([[L_\mu,L_\nu],L^\nu]+[[R_\mu,R_\nu],R^\nu]\right)\right], \\
S_{\mu\nu}&=&-g_{\mu\nu}\frac{\fpi^2}{2}Tr[P^2]
+\frac{1}{4\qpar^2}[g_{\mu\nu}h^{\alpha}_{\alpha}-h_{\mu\nu}]
\eer
where 
\ber
L_\mu&=&U^{\dag}\partial_\mu U, \;\;\;\;\; R_\mu = U\partial_\mu U^{\dag},\\
P&=&Q-U^{\dag}QU, \;\;\;\;\; h_{\mu\nu} = Tr[PL_\mu PL_\nu - P^2 L_\mu L_\nu]. 
\eer

The Lagrangian ${\cal L}_0$ has the classical solution for the configuration
called the hedgehog
\beq
U_h=exp[i\mbox{\boldmath{$\tau$}}\cdot\hat{\bf r} F(r)].
\eeq
The profile function $F(r)$ satisfy the following nonlinear equation
\ber
\frac{d^2 F}{d\tilde{r}^2}(\tilde{r}^2+2 sin^2F)
+2\frac{d F}{d\tilde{r}}\tilde{r}+(\frac{d F}{d\tilde{r}})^2sin(2F)
-sin(2F)(1+\frac{sin^2 F}{\tilde{r}^2}) \nonumber \\
-(\frac{m_\pi}{\fpi\qpar})^2 sin(F) \tilde{r}^2
=0 \label{proeq}
\eer
where $\tilde{r}=\qpar\fpi r$. For the topological charge $B=1$, the boundary
condition is
\beq
F(0)=\pi, \;\;\; F(\infty)=0.
\eeq

\section{Notations and generalized polarizabilities}
In this article  the virtual Compton scattering refers to the reaction
\beq
\gamma^*+p \rightarrow \gamma+p
\eeq
where $\gamma^*$, $\gamma$ and $p$ are respectively a space-like
virtual photon, a real photon and a proton. The momentum of 
the virtual (real) photon is represented by $q$ ($q'$). 

According to the low energy theorem (LET) the leading order terms 
of the scattering amplitude in an expansion of $q'_0$ is completely 
determined by the Born amplitude. And it depends only on the ground
state properties of the nucleon.  Guichon {\it et al.}~\cite{exp4} extended LET to include the higher order terms in terms of a multipole expansion.
The higher order terms rely on the excitation of the nucleon 
and are determined by the non-Born amplitude. The same authors developed
an appropriate formalism to describe the non-Born amplitude and defined
the so-called generalized polarizabilities (GPS), which are functions of $q^2$.
The GPS include the electric and the magnetic polarizabilities defined
through the real Compton scattering amplitude or the static energy shifts in
the presence of uniform electromagnetic fields.
We refer the details of the analysis of GPS to the reference~\cite{exp4}
and use their definitions only.
 
A generalized polarizability(GP) is defined as 
\beq
P^{(\rho' L',\rho L)S}(q)=\frac{1}{q'^{L'}q^{L}}
H^{(\rho' L',\rho L)S}_{NB}(q',q)\mid_{q'=0},
\;\;\;\;\; \rho',\rho=0\;\; or\;\; 1,
\eeq
where the reduced multipoles are defined by
\ber
H^{(\rho' L',\rho L)S}_{NB}(q',q)=
\frac{1}{2S+1}\sum_{\sigma'\sigma M'M}(-1)^{1/2+\sigma'+L+M}\nonumber \\
\times
\bra \frac{1}{2},-\sigma';\frac{1}{2},+\sigma\mid S s\ket
\bra L',M';L,-M\mid S s\ket
H^{\rho' L'M',\rho L M}_{NB}(q'\sigma',q\sigma).
\eer
The expression for the multipole is
\ber
H^{\rho' L'M',\rho L M}_{NB}(q'\sigma',q\sigma)=
\int d\hat{\bf q}\int d\hat{\bf q'}
V_{\mu}^{*}(\rho' L'M',\hat{\bf q'})
H^{\mu\nu}_{NB}({\bf q'}\sigma',{\bf q}\sigma)
V_{\nu}    (\rho  L M ,\hat{\bf q }),
\eer
where the $H^{\mu\nu}_{NB}(\bf{q'}\sigma',\bf{q}\sigma)$ is the non-Born
VCS amplitude and the $V_{\mu}(\rho  L M ,\bf{\hat{q}})$
$(\rho=0,\ldots, 3)$
are the complete 4-basis vectors defined as

\ber 
\label{eq:v01def} \\
&V^\mu (0LM,\hat{\bf q})= \left[ \begin{array} {c} 
Y^{L}_{M}(\hat{\bf q}) \\ {\bf 0}
\end{array}\right],&
V^\mu (1LM,\hat{\bf q})=
\left[\begin{array}{c}
0 \\ \frac{{\bf L_q}Y^{L}_{M}(\hat{\bf q})}{\sqrt{L(L+1)}}
\end{array}\right],\\
&V^\mu (2LM,\hat{\bf q})=
\left[\begin{array}{c}
0 \\ -i\hat{\bf q}\times\frac{{\bf L_q}Y^{L}_{M}(\hat{\bf q})}{\sqrt{L(L+1)}}
\end{array}\right],&
V^\mu (3LM,\hat{\bf q})=
\left[\begin{array}{c}
0 \\ \hat{\bf q}Y^{L}_{M}
\end{array}\right].
\eer

To describe the low energy behavior of the VCS, one needs 10 GP's
and the charge conjugation symmetry reduces the number to 6.
Among those, the familiar electric and the magnetic polarizabilites are
\beq
\alpha(q)=-\sqrt{\frac{3}{2}}e^2 P^{(01,01)0}(q),\;\;\;
\beta(q)= -\sqrt{\frac{3}{8}}e^2 P^{(11,11)0}(q).
\eeq
Using the definitions given in Eq.(\ref{eq:v01def}) we obtain
\ber
\label{eq:alpha}
\alpha (q)=\frac{1}{2}\left(\frac{e}{4\pi}\right)^2\frac{1}{q'q}
\int d\hat{\bf q}\int d\hat{\bf q'}\sum_{\sigma= \pm 1/2}
\hat{q'}_i
H^{00}_{NB}({\bf q'}\sigma',{\bf q}\sigma)
\hat{q}_i 
\mid_{q'=0}, \\
\beta (q)=\frac{1}{2}\left(\frac{e}{4\pi}\right)^2\frac{1}{q'q}
\int d\hat{\bf q}\int d\hat{\bf q'}\sum_{\sigma= \pm 1/2}
\epsilon_{ila}\hat{q'}_l
H^{ab}_{NB}({\bf q'}\sigma',{\bf q}\sigma)
\epsilon_{imb}\hat{q}_m 
\mid_{q'=0},
\eer
where the integrations are over 
the spherical angles of ${\bf q}$ and ${\bf q'}$ 
and the amplitude is spin-averaged. 
Here we also give the formula of $\beta(q)$ for future use.

\section{Results and conclusions}
As is clear from the interaction (\ref{eq:lagt}), the $\alpha(q)$ will have two
contributions; one is the seagull term and the other the dispersive term.
The seagull term, $\alpha_s$, is coming from the Lagrangian quadratic 
in $A_\mu$, Eq.~(\ref{eq:lag2}). 
The dispersive term, $\alpha_d$, is coming from the second order perturbation
theory applied to the linear interaction in $A_\mu$, Eq.~(\ref{eq:lag1}).
 
The computation of the $\alpha_d$  needs the calculation of the transition
matrix elements between the nucleon and the negative parity excited states.
\beq
\alpha_d(q)=2\sqrt{\frac{3}{2}}\sum_X
\bra N\mid d(0)|X\ket\bra X\mid d(q)\mid N\ket\frac{1}{m_N-m_X},
\eeq
\beq
d(q)=\int d^3r z J_0({\bf r})\frac{3j_1(qr)}{qr},
\eeq
where $j_1$ is the 1-st order spherical Bessel function. 
The negative parity excited state, $X$, most contributing to the electric dipole matrix element   
$\bra X\mid d(q)\mid N\ket$ is the bound state of the ground state nucleon
and the quantum pion. 
There is a general belief that the $\alpha_d(0)$ is much smaller
than the $\alpha_s(0)$~\cite{dsmall,hyperon}. Thus we neglect the
dispersive term.

Using the Lagrangian (\ref{eq:lag2}) and the formula given 
in Eq.~(\ref{eq:alpha}) we obtain
\beq
\alpha(q)=e^2\frac{8\pi}{9}\int_{0}^{\infty}dr 
r^4 sin^{2}F[\fpi^2+\frac{1}{\qpar^2}(F'^2+\frac{sin^{2}F}{r^2})]
\frac{3j_1(qr)}{qr}.
\eeq
Therefore
\ber
\label{eq:alpha0}
\alpha(0)=
e^2\frac{8\pi}{9}\int_{0}^{\infty}dr 
r^4 sin^{2}F[\fpi^2+\frac{1}{\qpar^2}(F'^2+\frac{sin^{2}F}{r^2})],
\\
\label{eq:alphad0}
\frac{d\alpha(q)}{d(q^2)}\mid_{q=0}=
-e^2\frac{4\pi}{45}\int_{0}^{\infty}dr 
r^6 sin^{2}F[\fpi^2+\frac{1}{\qpar^2}(F'^2+\frac{sin^{2}F}{r^2})].
\eer

For the purpose of an illustration we have performed a numerical
calculation for the set of parameters of 
$\fpi=93\; MeV,\;\;\; \qpar=4.26$, which gives the magnetic moments of
proton and neutron as 3.2 and -2.5 $\mu_N$, respectively.
Our numerical result is
\beq
\alpha(0)=8.44\times 10^{-4}\;\; fm^3,
\;\;\;\;\;
\frac{d\alpha(q)}{d(q^2)}\mid_{q=0}=-10.6\times 10^{-4}\;\; fm^5.
\eeq
The value of the $\alpha(0)$ is comparable with the empirical value
\beq
\alpha(0)^{expt}=(10.9\pm 2.2\pm 1.4 )\times 10^{-4}\;\; fm^3.
\eeq
The experimental value of the $\frac{d\alpha(q)}{d(q^2)}\mid_{q=0}$ 
is not yet available.

\begin{figure}
\setlength{\epsfysize}{5in}
\centerline{\epsffile{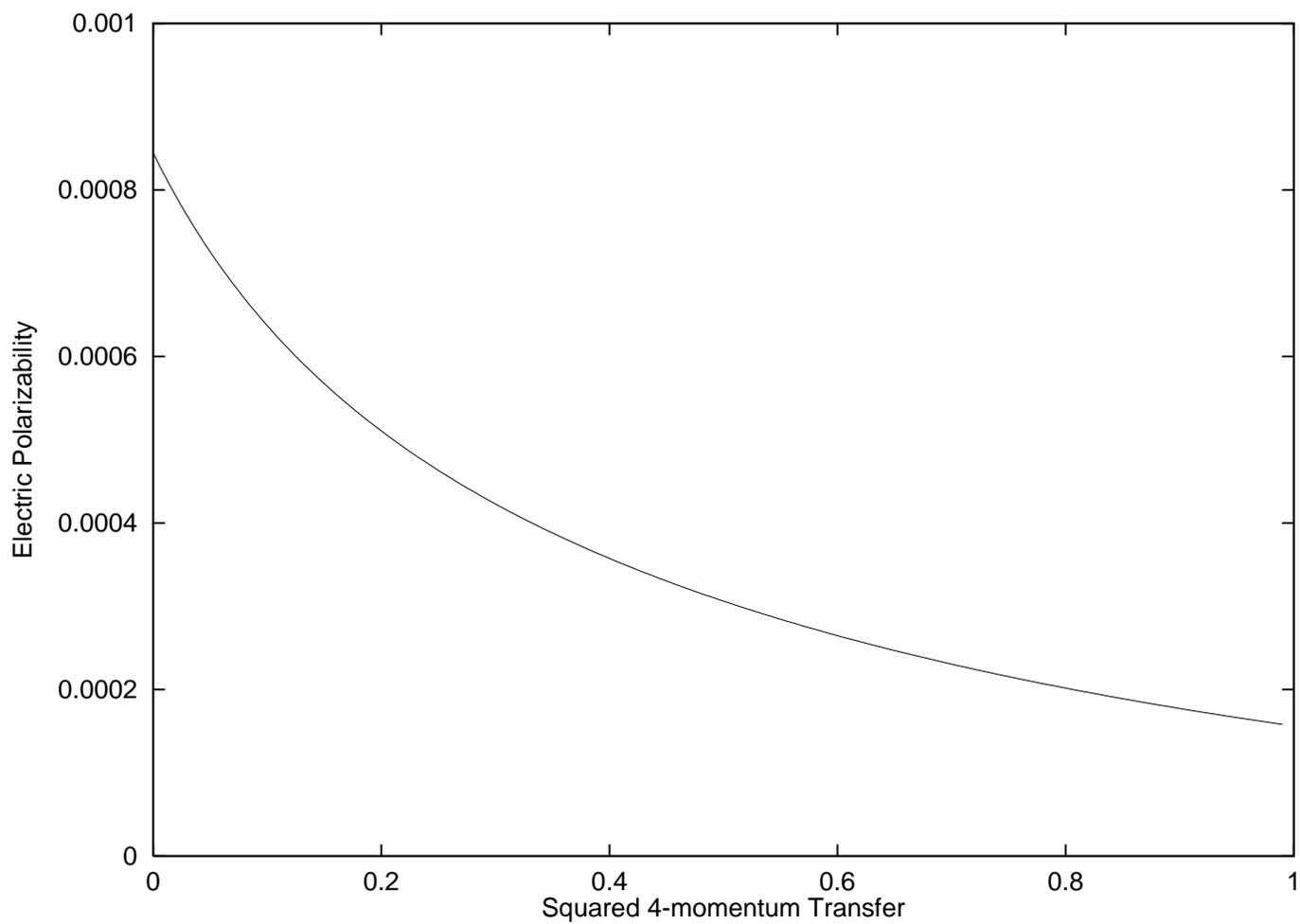}}
\vskip 2cm
\caption{The squared 4-momentum dependence of the electric polarizability.
The horizontal axis is in unit of $GeV^2$ and the vertical axis is in unit of
$fm^3$.}\label{fig}
\end{figure}

In Fig.~\ref{fig}, we plot the $q^2$ dependence of the generalized 
electric polarizability. The experiment is being performed but results are 
not yet 
available to compare with our plot. Our result is comparable
with the predictions of the linear sigma model~\cite{LSM}
and the constituent quark model~\cite{exp4}, the comparisons with which 
are made in the following in addition to the comparison with CHPT~\cite{CHPT1}.

Using the equation (\ref{proeq}) and the Goldberger-Treiman relation~\cite{GT}
 one can easily show that the function $F(r)$ behaves asymptotically as
\beq
F(r)\sim \frac{3}{8\pi}\frac{g_A}{\fpi^2}
\frac{1}{r^2}(1+m_\pi r)e^{-m_\pi r}.
\eeq
Thus 
$\alpha_s(0)$ and $\frac{d^2\alpha_s(q)}{dq^2}\mid_{q=0}$
diverge in the chiral limit since 
their integrands given in Eqs.~(\ref{eq:alpha0}) and (\ref{eq:alphad0})
do not asymptotically vanish.  
We pick up the diverging chiral behaviors of the quantities which read
\ber
\alpha(0)\rightarrow e^2\frac{5}{32\pi}\left(\frac{g_A}{\fpi}\right)^2
\frac{1}{m_\pi}, \\
\frac{d\alpha(q)}{d(q^2)}\mid_{q=0}\rightarrow
-e^2\frac{1}{64\pi}\left(\frac{g_A}{\fpi}\right)^2
\frac{1}{m_\pi^3}.
\eer
These agree with the results of the linear sigma model (LSM)~\cite{LSM} and 
the chiral perturbation theory (CHPT)~\cite{CHPT1}
up to a numerical factor. 
Our leading chiral terms are three times larger than theirs. 
The differences can be explained by
the contribution of the degenerate $\Delta$ states which were not included
in LSM and CHPT. 
More explicitly in the limit of $m_\pi\rightarrow 0$  
\beq
\alpha^{Skyrme}=[1+\frac{8}{9}
\left(\frac{g_{\pi N\Delta}}{g_{\pi NN}}\right)^2]
\alpha^{CHPT}
\eeq
where the ratio $\frac{g_{\pi N\Delta}}{g_{\pi NN}}$ is $\frac{3}{2}$ 
in the Skyrme model and the number $\frac{8}{9}$ is the characteristic of a
scalar-isoscalar operator. Numerically the value of the leading 
chiral term $\alpha^{Skyrme}$ ($\alpha^{CHPT}$) only 
overestimates (underestimates)
the electric polarizability. In the Skyrme model the full expression given
in Eq.~(\ref{eq:alpha}) is smaller than $\alpha^{Skyrme}$. In CHPT
the next order chiral terms include the contribution of the $\Delta$ state
with finite $N-\Delta$ splitting via a low energy constant~\cite{Cohen}.

Of course
the detailed analysis of the virtual Compton scattering described
by the generalized polarizabilities requires high quality data
which will be available in near future.  

\acknowledgements
We are thankful to Yongseok Oh for valuable discussions.
This work was supported by the Korean Science and Engineering
Foundation through the Center for Theoretical Physics
and by Korean Ministry of Education through contract number ISBR 96-2418. 

\end{document}